\documentclass[prl,aps,twocolumn,superscriptaddress,preprintnumbers,floatfix,longbibliography]{revtex4-1}
\pdfoutput=1
\usepackage{amsmath,amssymb,amsfonts,bm,graphicx,epsf,colordvi,bbm,pifont,siunitx,enumitem,hyperref,xcolor,multirow}

\DeclareRobustCommand{\Tab}[1]{Table~\ref{#1}}

\DeclareRobustCommand{\Fig}[1]{Fig.~\ref{#1}}

\DeclareRobustCommand{\Eq}[1]{Eq.~(\ref{#1})}

\DeclareRobustCommand{\Ref}[1]{Ref.~\cite{#1}}
\DeclareRobustCommand{\Refs}[1]{Refs.~\cite{#1}}

\newcommand{\x}{{\bf{x}}}
\providecommand{\href}[2]{#2}

\newcommand{\cmark}{\ding{51}}
\newcommand{\xmark}{\ding{55}}

\begin{document}

\title{Learning to Classify from Impure Samples with High-Dimensional Data}
\preprint{MIT--CTP 4968}

\author{Patrick T. Komiske}
\email{pkomiske@mit.edu}
\affiliation{Center for Theoretical Physics, Massachusetts Institute of Technology, Cambridge, MA 02139, USA}

\author{Eric M. Metodiev}
\email{metodiev@mit.edu}
\affiliation{Center for Theoretical Physics, Massachusetts Institute of Technology, Cambridge, MA 02139, USA}

\author{Benjamin Nachman}
\email{bpnachman@lbl.gov}
\affiliation{Physics Division, Lawrence Berkeley National Laboratory, Berkeley, CA 94720, USA}

\author{Matthew D. Schwartz}
\email{schwartz@physics.harvard.edu}
\affiliation{Department of Physics, Harvard University, Cambridge, MA 02138, USA}

\begin{abstract}
A persistent challenge in practical classification tasks is that labeled training sets are not always available.
In particle physics, this challenge is surmounted by the use of simulations. 
These simulations accurately reproduce most features of data, but cannot be trusted to capture all of the complex correlations exploitable by modern machine learning methods. 
Recent work in weakly supervised learning has shown that simple, low-dimensional classifiers can be trained using only the impure mixtures present in data. 
Here, we demonstrate that complex, high-dimensional classifiers can also be trained on impure mixtures using weak supervision techniques, with performance comparable to what could be achieved with pure samples.
Using weak supervision will therefore allow us to avoid relying exclusively on simulations for high-dimensional classification.
This work opens the door to a new regime whereby complex models are trained directly on data, providing direct access to probe the underlying physics. 
\end{abstract}

\maketitle

Data analysis methods at the Large Hadron Collider (LHC) rely heavily on simulations.
These simulations are generally excellent and allow us to explore the mapping between truth information (particles from collisions) and observables (reconstructed momenta and energies).
In particular, simulations let us train complex algorithms to extract the truth information from the observables. 
Machine learning methods trained on low-level inputs have been developed for collider physics~\cite{Larkoski:2017jix} to identify boosted $W/Z/$Higgs bosons~\cite{Cogan:2014oua,deOliveira:2015xxd,Baldi:2016fql,Barnard:2016qma,Louppe:2017ipp,Datta:2017rhs,Komiske:2017aww}, top quarks~\cite{Almeida:2015jua,Kasieczka:2017nvn,Pearkes:2017hku,Butter:2017cot,Egan:2017ojy}, $b$-quarks~\cite{CMS-DP-2017-005,ATL-PHYS-PUB-2017-003,Sirunyan:2017ezt}, and light quarks~\cite{Komiske:2016rsd,CMS-DP-2017-027,ATL-PHYS-PUB-2017-017,Bhimji:2017qvb}, for removing noise~\cite{Komiske:2017ubm}, and for emulating particle interactions with calorimeters~\cite{deOliveira:2017pjk,Paganini:2017hrr,Paganini:2017dwg}.
These new methods achieve excellent performance by exploiting subtle features of the simulations, which are presumed to be similar to the features in the data.
 Unfortunately, the simulations are known to be imperfect.
This is particularly true for subtle features in high-dimensions, as illustrated clearly for boosted $W$ bosons in Ref.~\cite{Barnard:2016qma} and by the need for non-negligible corrections (``scale factors'') to be applied to multivariate classifiers used by the current LHC experiments (see e.g.~\Refs{Chatrchyan:2012jua,CMS:2013kfa,Aad:2014gea,Khachatryan:2014vla,CMS:2014fya,Aad:2015ydr,Aad:2015rpa,Aad:2016pux}). 
Thus it is natural to question the performance of machine learning algorithms trained on simulations as we know that if a model is trained on unphysical artifacts, this is what the model will learn.
This objection certainly has merit, as the power of these methods for physics applications stems precisely from their ability to find features that we do not fully understand and cannot easily interpret.

Data-driven approaches avoid the pitfalls of relying on simulations in experimental analyses.
For simple observables, such as the invariant mass of a photon pair, a traditional experimental approach has been to perform sideband fits directly to the data.
This avoids relying on the simulation altogether.
Unfortunately, most of the sophisticated discrimination techniques developed in recent years use {\it full supervision}, where truth information is needed in order to train the classifier.
However, real data generally consist only of mixed samples without truth information, arising from underlying statistical or quantum mixtures of two classes (henceforth referred to as ``signal" and ``background").
Occasionally one can find a small region of phase space where the signal or background is pure, but these regions are generally sparsely populated and may not produce representative distributions.
Recent work on \emph{weak supervision}~\cite{hernandez2016weak} allows classifiers to be trained using only the information available from mixed samples.
Two weakly supervised paradigms tailored to physics applications are Learning  from Label Proportions (LLP)~\cite{Dery:2017fap} and Classification Without Labels (CWoLa)~\cite{Metodiev:2017vrx}.
\Ref{Dery:2017fap} considered the problem of discriminating the radiation pattern of quark from gluons ($q$/$g$) using three standard observables and showed how to achieve fully supervised discrimination power by using LLP with two samples of different but known quark fractions.
In \Ref{Metodiev:2017vrx}, it was shown that the proportions are not necessary for training since the likelihood ratio of the mixed samples is monotonically related to the signal/background likelihood ratio, the optimal binary classifier for signal vs. background.

One potential objection to the weak-learning demonstrations in \Refs{Dery:2017fap,Metodiev:2017vrx,Cohen:2017exh} is that the dimensionality of the inputs used is small.
Indeed, for a one-dimensional discriminant one can extract the exact pure distributions from mixed samples using the fractions.
It is not obvious that weak supervision will succeed when trained on high-dimensional inputs where the feature space may be sparsely populated.
Indeed, the most powerful modern methods are trained on high-dimensional, low-level inputs, where numerically approximating and weighting the probability distribution is completely intractable.  These deep learning techniques can expose subtle correlations in many dimensions which are also much harder to model than simple low-dimensional features.

In this paper, we demonstrate that weak supervision can approach the effectiveness of full supervision on complex models with high-dimensional inputs.
As a concrete illustration, we use an image representation to distinguish the radiation pattern from high energy quarks from gluons (``jet images"~\cite{Cogan:2014oua}).   Convolutional neural networks (CNNs) are applied to the quark and gluon jet images, where the dimensionality of the inputs is $\mathcal{O}(1000)$ and simulation mis-modeling issues are a challenge~\cite{Aad:2014gea,ATLAS-CONF-2016-034,CMS-PAS-JME-13-002,CMS-DP-2016-070,CMS-PAS-JME-16-003,ATL-PHYS-PUB-2017-009}.
We find that CWoLa more robustly generalizes to learning with high-dimensional inputs than LLP, with the latter requiring careful engineering choices to achieve comparable performance.
Though we use a particle physics problem as an example, the lessons about learning from data using mixtures of signal and background are applicable more broadly.

We begin by establishing some notation and formulating the problem.
Let $\x$ represent a vector of observables (\emph{features}) useful for discriminating two classes we call \emph{signal} ($S$) and \emph{background} ($B$).
For example, $\x$ might be the momenta of observed particles, calorimeter energy deposits, or a complete set of observables~\cite{Datta:2017rhs,Komiske:2017aww}.
In fully supervised learning, each training sample is assigned a truth label such as 1 for signal and 0 for background.
Then the fully supervised model is trained to predict the correct labels for each training example by minimizing a loss function.
For a sufficiently large training set, an appropriate model parameterization, and a suitable minimization procedure, the learned model should approach the optimal classifier defined by thresholding the likelihood ratio.

Data collected from a real detector do not come with signal/background labels.
Instead, one typically has two or more \emph{mixtures} $M_a$ of signal and background with different signal fractions $f_a$, such that the distribution of the features, $p_{M_a}(\x)$, is given by:
\begin{equation}
\label{eq:decomp}
p_{M_a}(\x) = f_a \,p_S(\x) + (1 - f_a)\, p_B(\x),
\end{equation}
where $p_S$ and $p_B$ are the signal and background distributions, respectively.
Weak supervision assumes {\it sample independence}, that Eq.~\ref{eq:decomp} holds with the same distributions $p_S(\x)$
and $p_B(\x)$ for all mixtures. 
Although in most situations sample independence does not hold perfectly (see e.g. \Ref{Gras:2017jty}), it is often a very good approximation (cf. Table~\ref{tab:sampledep} below).

LLP uses any fully supervised classification method and modifies the loss function to globally match the signal fraction predicted by the model on a batch of training samples to the known truth fractions $f_a$.
Breaking the training set into batches, normally done to parallelize training, takes on a new significance with LLP since the loss function is evaluated globally on each batch. 
The batch size, which for LLP we define as the number of samples drawn from each mixture during one update of the model, is a critical hyperparameter of LLP.

The loss functions we use for LLP differ from those in \Ref{Dery:2017fap}.
Analogous to the mean squared error (MSE) loss function for fully supervised (or CWoLa) training, we introduce the weak MSE (WMSE) loss for the LLP framework:
\begin{equation}
\label{eq:wmse}\ell_{\rm WMSE} = \sum_{a}  \left(f_a - \frac{1}{N}\sum_{i=1}^{N}h(\x_i)\right)^2,
\end{equation}
where $N$ is the batch size, $a$ indexes the mixed samples, and $h$ is the model.
Analogous to the crossentropy, we also introduce the weak cross entropy (WCE) loss:
\begin{equation}
\label{eq:wce}\ell_{\rm WCE} = \sum_{a} \text{CE}\left(f_a,\, \frac{1}{N} \sum_{i=1}^{N}h(\x_i)\right),
\end{equation}
where $\text{CE}(a, b) = - a \log b - (1 - a) \log (1 - b)$.
One caveat we discovered while exploring LLP is that the range of $h(\x)$ must be restricted to $[0,1]$, otherwise the model falls into trivial minima of the loss function. 
We also observe the effect of model outputs becoming effectively binary at 0 and 1, necessitating additional care to avoid numerical precision issues.

CWoLa classifies two mixtures, $M_1$ and $M_2$, from each other using any fully supervised classification method.
The resulting classifier is then used to directly distinguish the original signal and background processes.
Amazingly, the CWoLa classifier asymptotically (as the amount of training data increases) approaches an ideal classifier trained on pure samples~\cite{Metodiev:2017vrx,scott2013,Cranmer:2015bka}. 
CWoLa does not require that the fractions $f_a$ are known for training (the fractions on smaller test sets can be used to calibrate the classifier operating points).
The CWoLa framework has the nice property that as the samples approach complete purity ($f_1\to0,\,f_2\to1$) it smoothly approaches the fully supervised paradigm.
CWoLa presently only works with two mixtures; if more than two are available they can be pooled at the cost of diluting their purity.
The key features of CWoLa and LLP are compared in \Tab{tab:comparison}.
Note that no learning is possible with either method as $f_1\rightarrow f_2$.

\begin{table}[t]
\centering
\begin{tabular}{|l|ccc|}
\hline
& \includegraphics[scale=0.025]{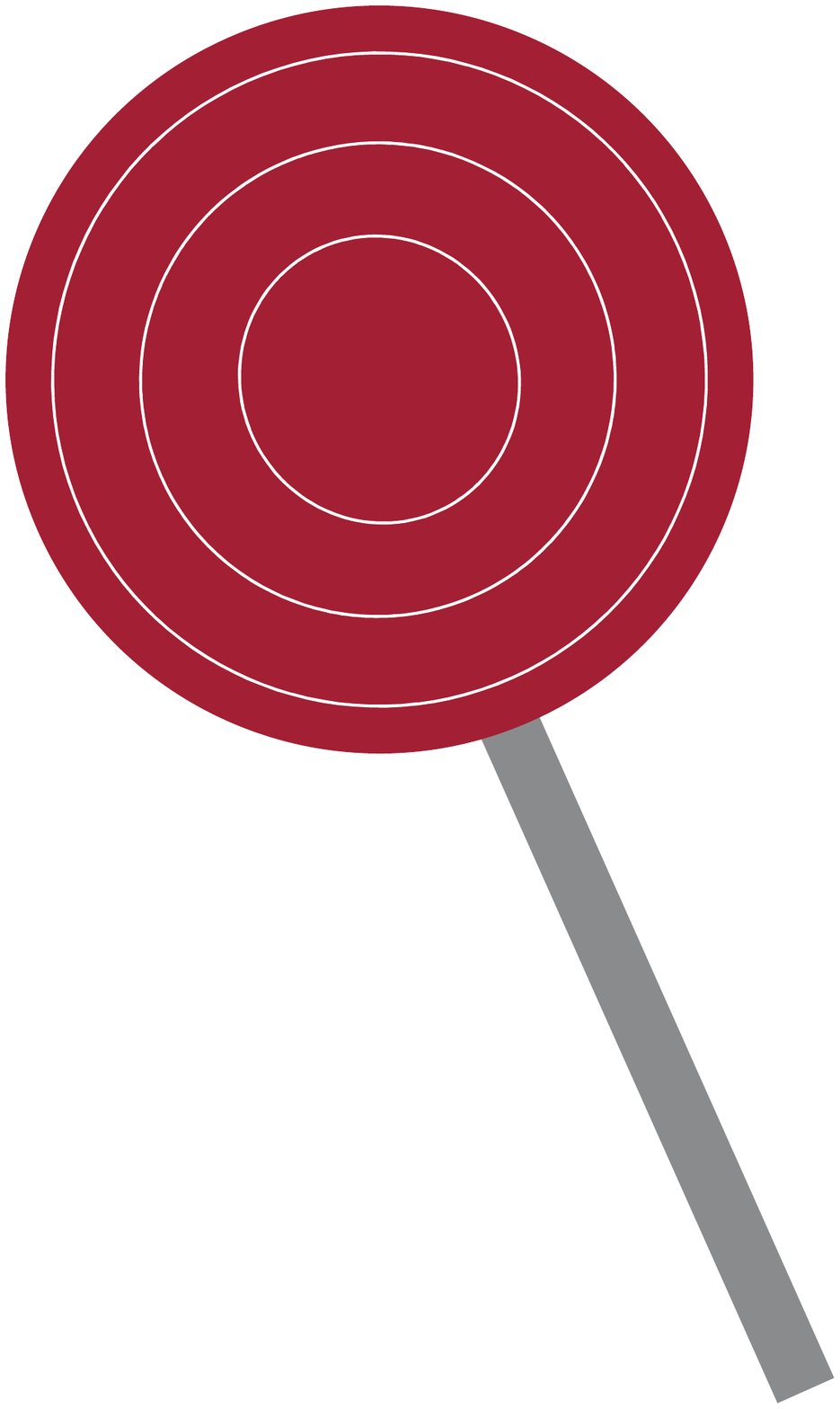} & \hspace{0.5mm} & \includegraphics[scale=0.025]{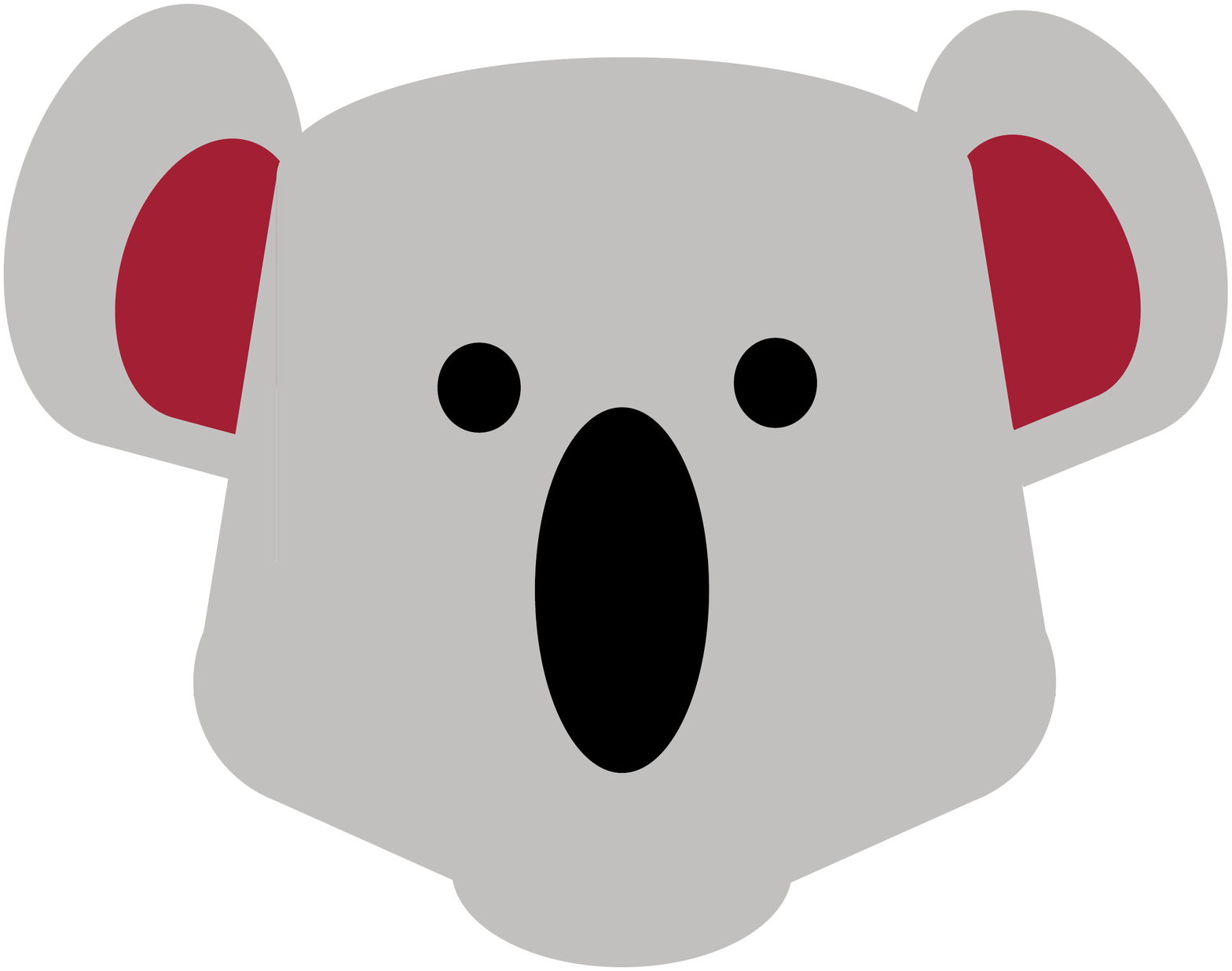} \\
\multicolumn{1}{|c|}{\bf Property} &{\bf LLP} & \hspace{0.5mm} & {\bf CWoLa}\\
\hline \hline
Compatible with any trainable model & \cmark & & \cmark \\
No training modifications needed & \xmark & & \cmark\\
Training does not need fractions & \xmark & & \cmark\\
Smooth limit to full supervision & \xmark & & \cmark \\ 
Works for $>2$ mixed samples & \cmark &  & \textbf{?} \\
\hline
\end{tabular}
\caption{The essential pros (\cmark), cons (\xmark), and open questions (\textbf{?}) of the CWoLa and LLP weak supervision paradigms.}
\label{tab:comparison}
\end{table}

To explore weak supervision methods with high-dimensional inputs, we simulate $Z+q/g$ events at $\sqrt{s}=13$ TeV using Pythia 8.226~\cite{Sjostrand:2007gs} and create artificially mixed samples with various quark (signal) fractions.
Jets with transverse momentum $p_T^{\text{jet}}\in[250,275]\text{ GeV}$ and rapidity $|y|\le2.0$ are obtained from final-state, non-neutrino particles clustered using the anti-$k_t$ algorithm~\cite{Cacciari:2008gp} with radius $R=0.4$ implemented in FastJet 3.3.0~\cite{Cacciari:2011ma}.
Single-channel, $33\times33$ jet images~\cite{Cogan:2014oua,deOliveira:2015xxd,Komiske:2016rsd} are constructed from a patch of the pseudorapidity-azimuth plane of size $0.8\times0.8$ centered on the jet, treating the particle $p_T$ values as pixel intensities.
The images are normalized so the sum of the pixels is 1 and standardized by subtracting the mean and dividing by the standard deviation of each pixel as calculated from the training set.

All instantiations and trainings of neural networks were performed with the python deep learning library Keras~\cite{keras} with the TensorFlow~\cite{tensorflow} backend.
A CNN architecture similar to that employed in \Ref{Komiske:2016rsd} was used: three 32-filter convolutional layers with filter sizes of $8\times 8$, $4\times 4$, and $4\times 4$ followed by a 128-unit dense layer.
Maxpooling of size $2\times2$ was performed after each convolutional layer with a stride length of 2.
The dropout rate was taken to be 0.1 for all layers.
Keras VarianceScaling initialization was used to initialize the weights of the convolutional layers. 
Due to numerical precision issues caused by the tendency of LLP to push outputs to 0 or 1, a softmax activation function was included as part of the loss function rather than the model output layer. 
Validation and test sets were used consisting each of 50k 50\%-50\% mixtures of quark and gluon jet images. 
Training was performed with the Adam algorithm~\cite{adam} with a learning rate of 0.001 and a validation performance patience of 10 epochs.
Each network was trained 10 times and the variation of the performance was used as a measure of the uncertainty.
Unless otherwise specified, the following are used by default: Exponential Linear Unit (ELU)~\cite{clevert2015fast} activation functions for all non-output layers, the CE loss function for CWoLa, and the WCE loss function for LLP.

The performance of a binary classifier can be captured by its receiver operating characteristic (ROC) curve.
To condense the classifier performance into a single number, we use the area under the ROC curve (AUC).
The AUC is also the probability that the classifier output is higher for signal than for background.
Random classifiers have $\text{AUC}=0.5$ and perfect classifiers have $\text{AUC}=1.0$.
We also confirmed that our conclusions are unchanged when using the background mistag rate at 50\% signal efficiency as a performance metric instead.

\begin{figure}[t]
\centering
\includegraphics[width=\columnwidth]{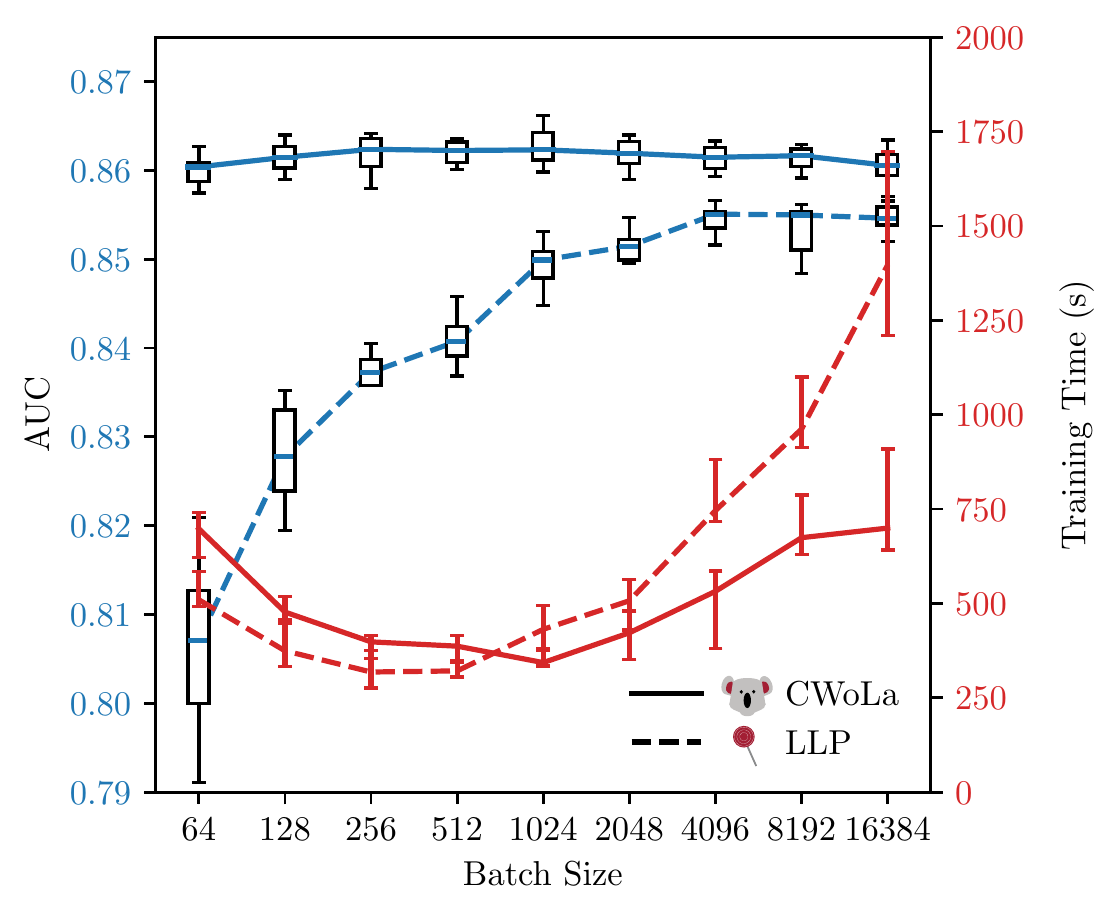}
\caption{The AUC and training time of CWoLa (solid) and LLP (dashed) as the batch size is varied. Training times are measured on an NVIDIA Tesla K80 GPU using CUDA 8.0, TensorFlow 1.4.1, and Keras 2.1.2.  AUC is a measure of classifier performance and is 1 for a perfect classifier and 0.5 for a completely random one.}
\label{fig:batchsweep} 
\end{figure}

As previously noted, the LLP paradigm works by matching the predicted fraction of signal events to the known fraction for multiple mixed samples.
In \Ref{Dery:2017fap}, the averaging took place over the entire mixed sample. 
Averaging over the entire training set at once is effectively impossible for high-dimensional inputs such as jet images because the graphics processing units (GPUs) that are needed to train the CNNs in a reasonable amount of time typically do not have enough memory to hold the entire training set at one time.
Hence, the ability to train with batches is highly desirable for using LLP with high-dimensional inputs.

There are many tradeoffs inherent with choosing the LLP batch size. 
Smaller batch sizes are susceptible to shot noise in the sense that the actual signal fraction on that batch may differ significantly from the fraction for the entire mixed sample, an effect which decreases as the batch size increases.
Smaller batch sizes result in longer training times per epoch (because the full parallelization capabilities of the GPU cannot be used) but often require fewer epochs to train.
Larger batch sizes have shorter training times per epoch but typically require more epochs to train.
For CWoLa, the batch size plays the same role as in full supervision, with the performance being largely insensitive to it but the total training time varying slightly.
These tradeoffs are captured in \Fig{fig:batchsweep}, which shows both the performance and training time for CWoLa and LLP models as the batch size is swept in powers of two from 64 to 16384, trained on two mixtures with $f_1 = 0.2$ and $f_2 = 0.8$.
The expected independence of CWoLa performance and the degradation of LLP performance for low batch sizes can clearly be seen.
The training time curves are concave with optimum batch sizes toward the middle of the swept region.
Based on this figure, we choose default batch sizes of $4000$ for LLP and 400 for CWoLa.

In order to explore a slightly more realistic scenario than artificially mixing samples from the same distribution of quarks and gluons, we generate $Z+\text{jet}$ and dijet events with the same generation parameters and cuts as described previously.
These ``naturally'' mixed samples have quark fractions $f_{Z+\text{jet}}=0.88$ and $f_{\text{dijets}}=0.37$.
The signal and background fractions have been systematically explored for these and many other processes in \Ref{Gallicchio:2011xc}.
As indicated by Table~\ref{tab:sampledep}, there is no significant difference in performance on the naturally mixed or artificially mixed samples.
Hence, artificially mixed samples are used in the rest of this study in order to evaluate weak supervision performance at different quark purities.
\begin{table}[t]
\centering
\begin{tabular}{| c |  l | l | c |}
\hline
\textbf{Learning} & \multicolumn{1}{c|}{\textbf{Sample}} & \multicolumn{1}{c|}{\textbf{AUC}} \\
\hline\hline
\multirow{2}{*}{CWoLa} & $Z$+jet vs. dijets & 0.8626 $\pm$ 0.0020 \\
 & Artificial $Z$ + $q$/$g$& 0.8621 $\pm$ 0.0019\\
\hline
\multirow{2}{*}{LLP} & $Z$+jet vs. dijets & 0.8544 $\pm$ 0.0019 \\
 & Artificial $Z$ + $q$/$g$& 0.8549 $\pm$ 0.0018 \\
\hline
\end{tabular}
\caption{AUCs for training with CWoLa and LLP on $Z+\text{jet}$ and dijet samples as well as on artificial mixtures of $Z+g$ and $Z+q$ samples. The error given is the interquartile range. There is no significant difference in classifier performance between the naturally mixed ($Z$+jet vs. dijets) samples and the artificially mixed ($Z + q/g$) samples with the same signal fractions.} 
\label{tab:sampledep}
\end{table}

\begin{figure}[t]
\centering
\includegraphics[width=\columnwidth]{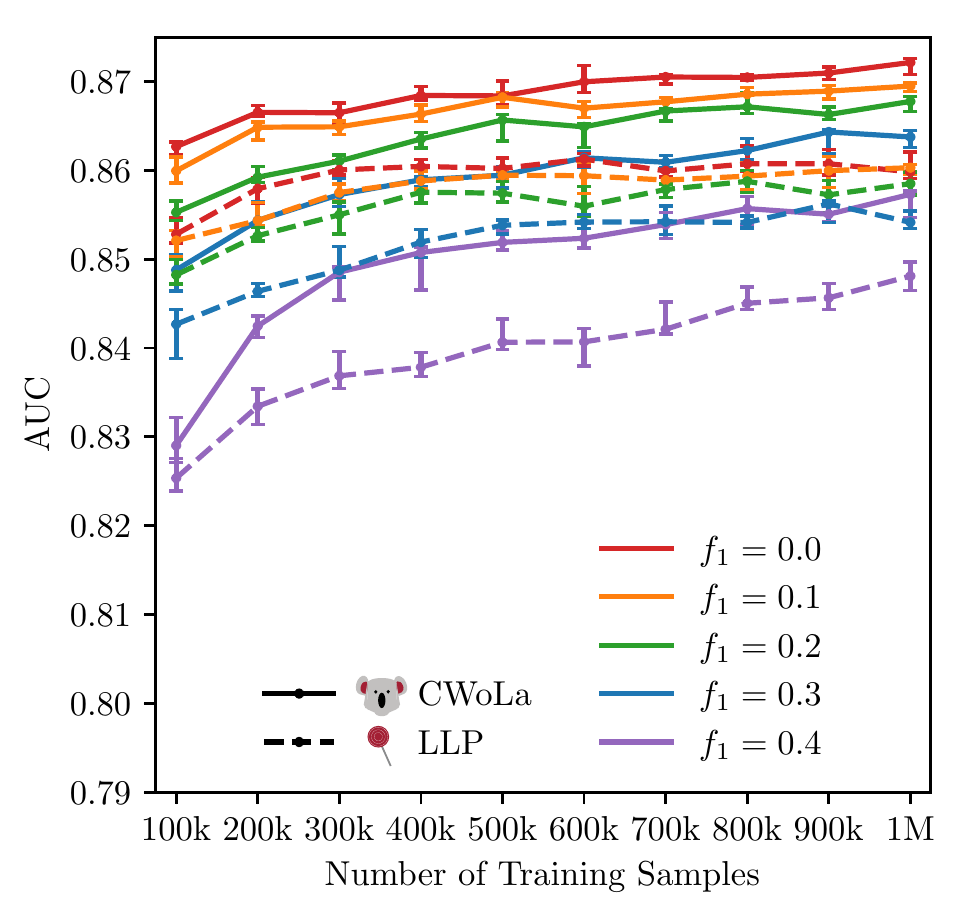}
\caption{Classifier performance (AUC) shown for both CWoLa (solid) and LLP (dashed) trained on two mixed samples with various signal fractions $f_1,\,1-f_1$ as the number of training data is varied between 100k and 1M. Each training is repeated 10 times and the $25^{\text{th}}$, $50^{\text{th}}$, and $75^{\text{th}}$ percentiles are shown. The $f_1 = 0.0$ CWoLa curve corresponds to full supervision. CWoLa outperforms LLP by this metric, though both methods work quite well.}
\label{fig:fracdatasweep}
\end{figure}

\Fig{fig:fracdatasweep} compares CWoLa and LLP performance for various quark/gluon purities as a function of the number of training samples.
Each network is trained using two samples, one with quark fraction $f_1$ and the other with quark fraction $f_2=1-f_1$.
Each point in the figure is the median of 10 independent network trainings and the error bars show the $25^{\text{th}}$ and $75^{\text{th}}$ percentiles.
Full supervision performance corresponds to CWoLa with $f_1=0$.
The most important takeaway from \Fig{fig:fracdatasweep} is that we have achieved good performance with both weak supervision methods over a large variety of sample purities and training sample sizes.
We also see that CWoLa consistently outperforms LLP and continues to get better as additional training samples are used, likely a result of the increasingly-populated feature space, whereas LLP performance tends to level off.
It should be noted that given the binary output nature of LLP models, classifiers trained in this way effectively come with a working point and sweeping the threshold to produce a ROC curve may not be ideal.
The purity/data tradeoff analysis of \Fig{fig:fracdatasweep} can provide valuable information for practical applications of weak supervision methods in physics, particularly in cases where more data can be acquired at the expense of worsening sample purity.

The sensitivity of LLP to different choices of loss function and activation function was examined.
We studied the choices of the symmetric squared loss of \Eq{eq:wmse} and the weak crossentropy loss of \Eq{eq:wce} with Rectified Linear Unit (ReLU)~\cite{nair2010rectified} and ELU activation functions.
We found a significant improvement in LLP classification performance in using ELU activations instead of ReLU activations, particularly at high signal efficiencies.
The choice of loss function was found to be less important than the choice of activation function, but minor improvements in AUC were observed with the WCE loss function over WMSE.
We also studied the dependence of CWoLa on the choice of activation function and found consistent performance between ELU and ReLU activations.
These results justify our default choices of ELU activation and WCE loss functions.
With the choice of ELU activation, LLP achieves almost the same performance to our CWoLa-trained network near the operating point with equal signal and background efficiencies.
We suspect this is a result of the tendency of LLP to output binary predictions (near 0 or 1) rather than a continuous output that can be easily thresholded.

Lastly, LLP has the potential advantage over the present implementation of CWoLa that it can naturally encompass multiple mixed samples with different purities. 
While in principle adding more samples should help, it is not obvious whether the network will effectively take advantage of them.
Indeed, we did not find significant improvement to LLP when adding additional samples with intermediate purities,
even after significant, dedicated architecture engineering.

In conclusion, we have shown that machine learning approaches using very high-dimensional inputs can be trained directly on mixtures of signal and background, and therefore on data.
This addresses one of the main objections to the use of modern machine learning in jet tagging: sensitivity to untrustworthy simulations.
We have implemented and tested weakly supervised learning with both LLP and CWoLa, finding that for the quark/gluon discrimination problem considered here CWoLa outperforms LLP and is less sensitive to particular hyperparameter choices.
We have developed a method for training LLP with high-dimensional inputs in batches and demonstrated that the batch size is a critical hyperparameter for both performance and training time. 
Given any fully supervised classifier, CWoLa works ``out-of-the-box'' whereas LLP requires additional engineering to achieve good performance and is generally harder to train.
Nonetheless, the success in using both of these weak supervision approaches on high-dimensional data is encouraging for the future of modern machine learning techniques in particle physics and beyond.

\acknowledgments

The authors would like to thank Lucio Dery and Francesco Rubbo for collaboration in the initial stages of this work.
We are grateful to Jesse Thaler for helpful discussions.
PTK and EMM would like to thank the MIT Physics Department for its support. 
Computations for this paper were performed on the Odyssey cluster supported by the FAS Division of Science, Research Computing Group at Harvard University.
This work was supported by the Office of Science of the U.S.~Department of Energy (DOE) under contracts DE-AC02-05CH11231 and DE-SC0013607, the DOE Office of Nuclear Physics under contract DE-SC0011090, and the DOE Office of High Energy Physics under contract DE-SC0012567.
Cloud computing resources were provided through a Microsoft Azure for Research award.
Additional support was provided by the Harvard Data Science Initiative.

\bibliography{highweaklett}

\end{document}